\documentclass[aps,twocolumn,pra]{revtex4}
\usepackage{graphicx}
\usepackage{amssymb}
\usepackage{bm}% bold math
\usepackage[latin1]{inputenc}

\begin{document}

\title{\bf Ordered Quantization and the Ehrenfest Time Scale}

\author{R.M. Angelo, L. Sanz and K. Furuya}

\affiliation{Instituto de F\'{\i}sica `Gleb Wataghin',CP 6165, Universidade
Estadual de Campinas, Unicamp\\ 13083-970, Campinas, SP, Brasil}

\begin{abstract}
We propose a prescription to quantize classical monomials in terms
of symmetric and ordered expansions of
non-commuting operators of a bosonic theory. As a direct application of 
such quantization rules, we quantize a classically time evolved function
 $\mathcal{O}(q,p,t)$, and calculate its expectation value in coherent 
states. The result can be expressed in terms of the application
of a classical operator which performs a {\em Gaussian smoothing} of the 
original function $\mathcal{O}$ evaluated at the center of the coherent 
state. This scheme produces a natural semi-classical expansion for the
quantum expectation values at a short time scale. Moreover, since the 
classical Liouville evolution of a Gaussian probability density gives the
{\bf same} form for the classical statistical mean value,
we can calculate the first order correction in $\hbar$ entirely
from the associated classical time evolved function. This allows us to 
write a general expression for the Ehrenfest time in terms of the 
departure of the centroid of the quantum distribution from the classical 
trajectory provided we start with an initially coherent state for 
each subsystem. In order to illustrate this approach, we have calculated 
analytically the Ehrenfest time of a model with $N$ coupled non-linear 
oscillators with non-linearity of even order.\\

\noindent
Pacs numbers: 03.65.Sq, 05.45.-a, 05.20.Gg
\end{abstract}
\maketitle
\newpage
%\setcounter{page}{2}
%%%%%%%%%%%%%%%%%%%%
\section{Introduction}
%%%%%%%%%%%%%%%%%%%

Since the earliest days of the quantum theory, the investigation of the
differences between the probabilistic concepts in classical Liouville
and quantum dynamics has been an important issue. There has been many
studies in the last two decades concerning the semi-classical regimen
of systems whose classical counterparts exhibits chaos \cite{1}.
The question of estimating how long the classical and quantum evolution
stay close has been one of the main questions in
semi-classical analysis. For classically chaotic flows, there was a
conjecture \cite{zaslavsky81} which is now rigorously proved 
\cite{hagerdorn00}, that the {\em break time} or Ehrenfest time ($t_E$) 
diverges logarithmically with $\hbar$.
In the classically regular flow, it was suggested in \cite{lai} that the 
behavior of $t_E$ with $\hbar$ is algebraic ($\hbar^{-\delta}$
), but no universal nature of such behavior has been shown yet.

According to the famous Ehrenfest's work \cite{ehrenfest}, for
quantum states which are localized enough, the time variation of the mean
quantum momentum must be equal to the local force. This statement
is exact for quadratic Hamiltonians, but its validity is
restricted to a short time scale, the Ehrenfest scale, for more
general non-linear systems. Mathematically, the Ehrenfest theorem
allows us to write $\langle \mathcal{O}(\hat{X})
\rangle=\mathcal{O}(\langle\hat{X}\rangle)$ for times smaller than 
the Ehrenfest time. 
In this situation, the initial dynamics is described essentially by a 
mean field approximation, where we have a localized packet obeying 
classical equations of motion. Then, it is reasonable to expect a rather 
good agreement between quantum and classical Liouvillian centroids and
the classical trajectory. In fact, this scenario have already been
reported in the literature \cite{ballentine98}.

In this work, we propose a simple analytical scheme to calculate
the Ehrenfest time for integrable systems. Our starting point
is to propose a classical Liouvillian operator that makes explicit the
symmetric form of the usual quantization rules. Using such an
operator, we are able to take a general
classical function $\mathcal{O}$ which expresses the time evolution of
a physical quantity, quantize it, and evaluate its expectation value in
coherent states at each time. The final result, which is shown to be
analytically identical to the statistical average calculated through
the classical Liouville formalism, is written in terms of a certain
differential operator acting onto the original classical function.
Since this is just a compact form of expressing
the action of the corresponding power series in $\hbar$,
our recipe automatically leads to a semi-classical expansion around
the time evolved classical function. Notice that the equality between
quantum and statistical centroids guarantees that we are working within
a classical time scale. This allows us to define mathematically the
Ehrenfest scale using just the Liouvillian Gaussian average without
solving the quantum dynamical problem.

 As an example, we present an explicit calculation 
of the Ehrenfest time for $N$-coupled non-linear oscillators, with 
nonlinearity of order $2k$, an even integer number. This model to 
which we will address the quantum-classical departure issue can be 
associated to several nonlinearly interacting fields via Kerr-type 
\cite{milburn,kitagawa} and cross-Kerr type interaction \cite{perinova} 
known to be relevant in quantum optics \cite{scully}, and also for two 
quantized vibrational modes of a single trapped ion \cite{steinbach}. 
It is an integrable model where the role of nonlinearity can be
studied analytically for several quantities, and, in particular, for
$N=k=2$ the quantum-classical {\em break time} has been determined 
\cite{sanz} based on physical properties of the exact quantum states. 
Although the emphasis of studying the break times in the 
literature has been on `chaotic states' \cite{ballentine,karkuszewski}, 
here we are concerned with integrable cases, where we are able to
derive an analytical expression for the Ehrenfest time.

%%%%%%%%%%%%%%%%%%%%%%%%%%%%%%%%%%%%%%%%%%%%%%%%%
\section{Ordered Quantization}
%%%%%%%%%%%%%%%%%%%%%%%%%%%%%%%%%%%%%%%%%%%%%%%%%
%%%%%%%%%%%%%%%%%%%%%%%%%%%%%%%%%%%%%%%%%%%%%%%
\subsection{Definition}
%%%%%%%%%%%%%%%%%%%%%%%%%%%%%%%%%%%%%%%%%%%%%%%%%%

Let us start by presenting a convenient quantization
scheme for a single degree of freedom system, which will
be easily generalized to degree $N$. Consider two
non-commuting operators $\hat{A}$ and $\hat{B}$ whose commutator
is a $c$-number, denoted by $c$ ($[\hat{A}, \hat{B}]=c$). Then, a
given classical monomial $a^nb^m$, where $a$ and $b$ are 
canonically conjugated classical variables \footnote{we 
are considering the cases where $a$ and $b$ are associated to the
operators $\hat{A}$ and $\hat{B}$ trough the coherent state mean
values, e.g., $\langle \alpha_o|\hat{A}|\alpha_o \rangle$. The
canonical pairs $(\hat{a},\hat{a}^{\dag})$ and $(\hat{Q},\hat{P})$
are both contempleted in our notation.}, 
will be quantized in a {\em symmetric} and {\em ordered} form, through 
the prescriptions:
\begin{eqnarray}
a^n b^m \rightarrow {\mathcal S}_{\hat{A},\hat{B}}\left(\hat{A}^n \hat{B}^m
\right)\,,\, \label{SAB}
\end{eqnarray}
where the super-operator ${\mathcal S}_{\hat{A},\hat{B}}$ is given by
\begin{eqnarray}
{\mathcal S}_{\hat{A},\hat{B}}=e^{-\frac{1}{2}[\hat{A},\hat{B}]
\partial_{\hat{A}}\partial_{\hat{B}}}=
\sum\limits_{k=0}^{\infty}\frac{\left(-\frac{1}{2}
[\hat{A},\hat{B}]\right)^k}{k!}\partial^k_{\hat{A}}
\partial^k_{\hat{B}}.
\label{ordops}
\end{eqnarray}
The above index $(\hat{A},\hat{B})$ corresponds to the ordering with 
$\hat{A}$ on the left and $\hat{B}$ on the right. 
Since $[\partial_{\hat{A}},\partial_{\hat{B}}]=[\partial_{\hat{A}},\hat{B}]
=[\hat{A},\partial_{\hat{B}}]=0$,
$\mathcal{S}_{\hat{A},\hat{B}}$ and $\mathcal{S}_{\hat{B},\hat{A}}$ are 
{\em classical} unitary differential super-operators, satisfying $\mathcal{S}_
{\hat{A},\hat{B}} \mathcal{S}_{\hat{B},\hat{A}} = 
\mathcal{S}_{\hat{B},\hat{A}}\mathcal{S}_{\hat{A},\hat{B}} = 1$. The 
prescription defined in expression (\ref{SAB}) leads to different 
orderings (depending on which variable is chosen to be $a$ or $b$) for 
the same original classical function and, therefore, the associated 
symmetric operators must be the same. From this consideration, we deduce 
the following ordering formulas:
\begin{equation}
\begin{array}{l}
\hat{A}^n\hat{B}^m=\mathcal{S}_{\hat{B},\hat{A}}^2\hat{B}^m\hat{A}^n =
e^{+[\hat{A},\hat{B}]\partial_{\hat{A}}
\partial_{\hat{B}}}\hat{B}^m\hat{A}^n, \\ \\
\hat{B}^m\hat{A}^n=\mathcal{S}_{\hat{A},\hat{B}}^2\hat{A}^n\hat{B}^m =
e^{-[\hat{A},\hat{B}]\partial_{\hat{A}}
\partial_{\hat{B}}}\hat{A}^n\hat{B}^m.
\end{array} \label{ord}
\end{equation}
Using a classical displacement operator,
$e^{a\partial_x}f(x)=f(x+a)$, which can be used to write
$e^{-\frac{c}{2}\partial_{\hat{A}}\partial_{\hat{B}}}\hat{A}^n\hat{B}^m=
(\hat{A}-\frac{c}{2}\partial_{\hat{B}})^n\hat{B}^m$, one can show
that these results reproduce those in the textbooks (see e.g.
Louisell \cite{louisell}). A simple example shows that our
quantization scheme leads to the usual quantization rules, but in
an automatically ordered form. Consider the following product of classical
canonical phase space variables: $q^2p$.
According to the usual rules, this is transformed in a {\em totally symmetric 
operator} \cite{cahill} of the form
$\frac{1}{3}(\hat{Q}^2
\hat{P}+\hat{P}\hat{Q}²+\hat{Q}\hat{P}\hat{Q})$; and using the
commutation relation $[\hat{Q},\hat{P}]=\imath\hbar$, 
re-written as $\hat{Q}²\hat{P}-\imath \hbar\hat{Q}$. 
But this is exactly the result produced by the expression in (\ref{SAB}) 
with the choice $(a,b)=(q,p)$,  
just by making some few derivatives.
We finally note that the super-operators
$\mathcal{S}_{\hat{A},\hat{B}}$ has already appeared in the literature 
in the case of the canonical phase space operators
$(\hat{A},\hat{B})=(\hat{Q},\hat{P})$, in slightly different
contexts \cite{royer91}.
%%%%%%%%%%%%%%%%%%%%%%%%%%%%%%%%
\subsection{Normal Ordering in Bosonic Operators}
%%%%%%%%%%%%%%%%%%%%%%%%%%%%%%%%
As an immediate application of the formulas presented above, we will 
express the totally symmetric ordered expression 
$\mathcal{S}_{\hat{Q},\hat{P}}\hat{Q}^n\hat{P}^m$ in a form more suitable for
our purpose of taking expectation values in the Weyl-Heisenberg 
coherent states. We start by expressing such a polynomial in terms of 
bosonic creation and annihilation operators
\begin{eqnarray}
\mathcal{S}_{\hat{Q},\hat{P}}\hat{Q}^n\hat{P}^m&=&
e^{-\frac{\imath\hbar}{2}\partial_{\hat{Q}}\partial_{\hat{P}}}
\hat{Q}^n\hat{P}^m \label{sqp} \\
&=&z_{n,m}
e^{\frac{1}{4}\left(\partial_{\hat{a}}^2-\partial_{\hat{a}^{\dag}}^2
\right)}
\left(\hat{a}+\hat{a}^{\dag}\right)^n\left(\hat{a}-\hat{a}^{\dag}
\right)^m, \nonumber
\end{eqnarray}
where $z_{nm}=(-\imath)^m \left(\sqrt{\hbar/2}\right)^{n+m}$. Now,
using the fact that
\begin{eqnarray}
\left(\hat{a}\pm\hat{a}^{\dag} \right)^n
&=&\sum\limits_{k=0}^{n}\left(\pm1 \right)^{n-k}{n\choose
k}e^{\frac{1}{2}
\partial_{\hat{a}}\partial_{\hat{a}^{\dag}}}\left(\hat{a}^{\dag\,n-k}
\hat{a}^{k}\right), \label{a+a*}
\end{eqnarray}
we can re-express (\ref{sqp}) as
\begin{eqnarray}
&\mathcal{S}_{\hat{Q},\hat{P}}\hat{Q}^n\hat{P}^m=
\sum\limits_{k=0}^{n}\sum\limits_{l=0}^{m}(-1)^{m-l}z_{n,m}{n\choose
k}{m\choose l} \times
\nonumber \\
&\times\,
e^{\frac{1}{4}\left(\partial_{\hat{a}}^2-\partial_{\hat{a}^
{\dag}}^2
\right)}e^{\frac{1}{2}\partial_{\hat{a}_2}\partial_{\hat{a}^
{\dag}_1}}e^{\frac{1}{2}\partial_{\hat{a}_4}\partial_{\hat{a}^{\dag}_3}}
\left(\hat{a}^{\dag\,n-k}_1\hat{a}^{k}_2\hat{a}^{\dag\,m-l}_3\hat{a}^{l}_4
\right),
\end{eqnarray}
with the sub-indexes that we introduced, to indicate where the 
action of the differentiation should take place. At the end of 
calculation we must erase all these sub-indexes. Using the relation 
(\ref{ord}), we get
\begin{eqnarray}
&\mathcal{S}_{\hat{Q},\hat{P}}\hat{Q}^n\hat{P}^m=
\sum\limits_{k=0}^{n}\sum\limits_{l=0}^{m}(-1)^{m-l}z_{n,m}{n\choose
k}{m\choose l}
\times \label{sa*a}\\
& \times\,
e^{\frac{1}{4}\left(\partial_{\hat{a}}^2-\partial_{\hat{a}^
{\dag}}^2 \right)}
e^{\frac{1}{2}\partial_{\hat{a}_2}\partial_{\hat{a}^{\dag}_1}}
e^{\frac{1}{2}\partial_{\hat{a}_4}\partial_{\hat{a}^{\dag}_3}}e^{\partial_{
\hat{a}_2}\partial_{\hat{a}^{\dag}_3}} \left(\hat{a}^{\dag\,n-k}_1
\hat{a}^{\dag\,m-l}_3\hat{a}^{l}_2  \hat{a}^{k}_4\right).
\nonumber
\end{eqnarray}
This is the normal ordered expression (in the
creation and annihilation operators) for the original 
$\hat{Q}\hat{P}$-ordered monomial. Now, it is a simple matter to 
calculate its expectation value.
%%%%%%%%%%%%%%%%%%%%%%%%%%%%%%%%%%%%%%%%%%%%%%%%%%%%
\subsection{Coherent States Representation}
Since we are interested in the connection between quantum and
classical mechanics, the coherent state basis appears as the most 
appropriated one. In particular, it will be of particular interest for 
us to evaluate the expectation value in coherent states of some operator 
products like the one treated in previous section. Then, we first calculate
the matrix elements in the coherent state basis of the operator function
given in  Eq.(\ref{sa*a})
\begin{eqnarray}
\frac{\langle\alpha_1|\mathcal{S}_{\hat{Q},\hat{P}}\hat{Q}^m\hat{P}^m
|\alpha_2\rangle}{\langle\alpha_1|\alpha_2\rangle}
=z_{n,m}e^{\frac{1}{4}\left(\partial_{\alpha_2}^2-\partial_{\alpha_1^*}^2
\right)}\left\{ e^{\partial_{a}\partial_{d}}
e^{\frac{1}{2}\partial_{a}\partial_{b}}\times \right. \nonumber \\
\times \left. e^{\frac{1}{2}\partial_{c}\partial_{d}}
\left[\sqrt{\frac{\hbar}{2}}(a+b)\right]^n
\left[\sqrt{\frac{\hbar}{2}} \frac{(c-d)}{\imath}
\right]^m\right\}_{{a=c=\alpha_2\atop b=d=\alpha^*_1}}.
\label{f12}
\end{eqnarray}
Setting now $\alpha_1=\alpha_2=\alpha_0=\frac{q_0+\imath
p_0}{\sqrt{2\hbar}}$ and performing adequate variable
transformations, we finally get
\begin{eqnarray}
&\langle\alpha_0|\mathcal{S}_{\hat{Q},\hat{P}}\hat{Q}^n\hat{P}^m
|\alpha_0\rangle=e^{\frac{\hbar}{4}\nabla_0^2}q_0^n p_0^m,& \label{Oqp}
\end{eqnarray}
where
\begin{equation} \begin{array}{c}
\nabla_0²=\partial_{q_0}²+\partial_{p_0}², \\ \\
(q_0,p_0)=\left(\langle\alpha_0|\hat{Q}|\alpha_0\rangle,\langle\alpha_0|
\hat{P}|\alpha_0\rangle\right).
\end{array} \end{equation}
The expectation value given in (\ref{Oqp}) must be calculated
through the series expansion of the classical operator
$e^{\frac{\hbar}{4}\nabla_0^2}$, which gives a natural expansion in powers 
of $\hbar$, showing its semi-classical nature. 
In fact, in the classical limit $\hbar\rightarrow 0$, the quantum 
expectation value of the operator function reduces to a purely classical 
function calculated at the center of the coherent packet located at 
$(q_0,p_0)$.

Another interesting relation can be obtained from a similar
calculation
\begin{equation} \begin{array}{c}
\langle \hat{Q}^n\hat{P}^m\rangle=e^{\frac{\hbar}{4}\nabla_0^2}
\left(e^{\frac{\imath\hbar}{2}\partial_{q_0}\partial_{p_0}}q_0^n p_0^m
\right),
\end{array} \label{mv1}
\end{equation}
which implies that
\begin{equation} \begin{array}{c}
\frac{1}{\imath \hbar}\langle [\hat{Q}^n,\hat{P}^m]\rangle=
e^{\frac{\hbar}{4}\nabla_0^2}
\left(\frac{\sin{\frac{\hbar}{2}\partial_{q_0}\partial_{p_0}}}
{\frac{\hbar}{2}}\right)q_0^n p_0^m.
\end{array} \end{equation}
It is important to note that the term within the parentheses in 
Eq.(\ref{mv1}) is exactly the Weyl transform of the operator 
$\hat{Q}^n\hat{P}^m$ \cite{royer91}. The extra operator factor 
$e^{\frac{\hbar}{2}\nabla_0^2}$ is what makes reference to the width 
of the coherent packet, as we shall see latter. These results also point 
out to the existence of an asymptotic classical limit ($\hbar \rightarrow 0$)
for such expectation values in coherent states.

%%%%%%%%%%%%%%%%%%%%%%%%%%%%%%%%%%%%%%%%%%%%%%%
\section{Short Time Quantization}
%%%%%%%%%%%%%%%%%%%%%%%%%%%%%%%%%%%%%%%%%%%%%%%
The usual quantization rules are defined in the Heisenberg
picture, where the Hamilton's equations solution
$q_{\mathcal{H}}(q,p,t)$ and $p_{\mathcal{H}}(q,p,t)$, are
transformed into Heisenberg operators
$\hat{Q}_H(\hat{Q},\hat{P},t)$ and $\hat{P}_H(\hat{Q},\hat{P},t)$, where
$\hat{Q}$ and $\hat{P}$ denote Schr\"odinger operators. On the
other hand, since the Heisenberg and Schr\"odinger pictures
coincide at $t=0$, any scheme of quantization based on the Schr\"odinger 
picture would describe reasonably the quantum world for very short times. 
However, as long as we are interested in analyzing the quantum operator 
evolution only during a classical (mean field) time scale, this would 
suffice.

In this context, consider a classically time evolved function
$\mathcal{O}$ that has the following well defined expansion
\begin{eqnarray}
\mathcal{O}(q,p,t)=\sum\limits_{n,m=0}^{\infty}c_{n,m}(t)q^np^m.
\end{eqnarray}
The coefficients $c_{n,m}(t)$ contain all the time dependence. Now,
applying the operator of ordered quantization $\mathcal{S}_{\hat{Q},
\hat{P}}$ and taking the average in coherent states we obtain
\begin{eqnarray}
&\langle\alpha_0|\hat{\mathcal{O}}(\hat{Q},\hat{P},t)|\alpha_0\rangle=
e^{\frac{\hbar}{4}\nabla_0^2}\mathcal{O}\left(q_0,p_0,t\right),&
\label{Oq0p0}
\end{eqnarray}
where we have used the result (\ref{Oqp}).
In order to estimate for how long the result (\ref{Oq0p0}) could be 
trusted, we present the calculation of the classical
statistical counterpart of the problem. In the classical Liouvillian 
formalism, the following mean value is defined in the phase space 
$(q,p)$
\begin{equation}
\left< \mathcal{O}(q_0,p_0,t)\right>=\int dq dp
\,\,\rho(q,p,t)\,\mathcal{O}(q,p,0),
\label{eq:mean}
\end{equation}
where ($q_0$,$p_0$) stands for the center of the initial distribution
$\rho(q,p,0)$. Then, since
$\rho(q,p,t)=e^{\mathcal{L}t}\rho(q,p,0)$
and
$\mathcal{O}(q,p,t)=
e^{-\mathcal{L}t}\mathcal{O}(q,p,0)$,
taking into account the fact that the volume in the phase space is
preserved, we can re-write the previous equation in a
Heisenberg-like form
\begin{eqnarray}
\left< \mathcal{O}(q_0,p_0,t)\right>=\int
dq dp\,\rho(q,p,0)\,\mathcal{O}(q,p,t).
\label{eq:mv}
\end{eqnarray}
Consider now a Gaussian initial distribution with width $\sigma$, and 
centered at the point $(q_0,p_0)$. Performing the variable transformations,
$(q-q_0)=x$ and $(p-p_0)=y$, and using the classical
 displacement operator again, we can rewrite Eq.(\ref{eq:mv}) as
\begin{eqnarray}
\langle \mathcal{O}\rangle =
\int dx dy \frac{e^{-\frac{x^2}{\sigma}}}{\sqrt{\pi\sigma}}
\frac{e^{-\frac{y^2}{\sigma}}}{\sqrt{\pi\sigma}}e^{x \partial_{q}}
e^{y \partial_{p}}\mathcal{O}(q,p,t).
\label{eq:stat}
\end{eqnarray}
Notice that the function $\mathcal{O}$ inside the integral no longer 
depends on the integration variables. Hence, by completing squares and
performing formally the integration, we finally obtain
\begin{eqnarray}
\langle \mathcal{O}(q_0,p_0,t)\rangle =
 e^{\frac{\sigma}{4}\nabla_0^2}\mathcal{O}(q_0,p_0,t)
\label{OEst}.
\end{eqnarray}
Now, we have proved that the effect of applying the operator
$e^{\frac{\sigma}{4}\nabla_0^2}$ is exactly that of smoothing
the function $\mathcal{O}$ through a Gaussian mean, where $\sigma$ 
is related to the width of the Gaussian distribution to be used in the
smoothing. Accordingly, this operator has already been used to express the
$Q$-function as the Gaussian smoothing of the Wigner function \cite{cessa}. 
It is important to notice that this result is {\bf exact} for Gaussian 
statistical averages, i.e., it was derived from
first principles without any approximation. The only implicit
assumption used in the calculation was the existence of the
derivatives in all orders for the function $\mathcal{O}$.

The comparison between expressions (\ref{Oq0p0}) and (\ref{OEst}),
with $\sigma=\hbar$, confirms the fact that our proposal of a
Schr\"odinger quantization for classical function should be adequate
either at a short time scale or in cases in which the classical
function depends linearly on the phase space coordinates $(q_0,p_0)$
(e.g. harmonic oscillator). The results (\ref{Oq0p0})
and (\ref{OEst}) can easily be extended for higher degrees of freedom.

At this point one might formulate a question. First, we interpret
the action of the operator $e^{\frac{\hbar}{4}\nabla_0^2}$ on any classical 
function, as an exact factorization of the effect of a Gaussian wave packet 
contribution. 
In fact, for completely localized distribution ($\hbar=0$) there are no
corrections. Then, one may ask: is it possible, in the particular case
of coherent separable initial states, to express the exact
quantum centroid, $\langle\hat{\mathbf{R}}(t)\rangle=
\langle\big(\hat{Q}(t),\hat{P}(t)\big)\rangle$
in terms of the smoothed form 
$e^{\frac{\hbar}{4}\nabla^2}\mathbf{r_c}(t)$, where 
$\mathbf{r_c}(t)=\big(q_c(t),p_c(t) \big)$ is a certain classical dynamical 
vector in phase space? The answer will be {\bf positive} if we are able to 
calculate the inverse of the Gaussian smoothing operation on the quantum 
centroid vector in the phase space
\begin{eqnarray}
\mathbf{r_c}(t)=e^{-\frac{\hbar}{4}\nabla_0^2}\langle\alpha_0|
\mathbf{\hat{R}}(t)|\alpha_0\rangle.
\label{cqt}
\end{eqnarray}
This function $\mathbf{r_c}(t)$ will then be a {\em coherent quantum trajectory},
in the sense that it will carry all quantum dynamical information possible
to put in a trajectory, except the contribution due to a Gaussian smoothing
effect. For the trivial case of $N$ non-interacting harmonic oscillators and 
the case of a bilinear coupling between two harmonic oscillators,
the quantum coherent and classical trajectories coincide 
( $ \mbox{\boldmath$($}q_c(t),p_c(t)  \mbox{\boldmath$)$}= 
\mbox{\boldmath$($}q_0(t),p_0(t)\mbox{\boldmath$)$}$ ), as expected.

%%%%%%%%%%%%%%%%%%%%%%%%%%%%%%%%%%%%%%%%%%%%
\section{The Ehrenfest Time}
%%%%%%%%%%%%%%%%%%%%%%%%%%%%%%%%%%%%%%%%%%%%
\subsection{Formal Definition}
%%%%%%%%%%%%%%%%%%%%%%%%%%%%%%%%%%%%%%%%%%%%
Now, we have all the necessary tools to undertake the problem of
estimating the Ehrenfest time in the case of an initially separable 
coherent Gaussian wave packet. Assuming that the break instant occurs 
when the first order correction in $\hbar$ becomes as important as the 
original vector, we apply the smoothing process to the phase space 
classical trajectory vector of a system with $N$ degrees of freedom 
$\mathbf{r}(t)= \mbox{\boldmath$($} q_1(t),p_1(t),\cdots,q_N(t),p_N(t)
\mbox{\boldmath$)$}$. Expanding the smoothing operator up to the first
order in $\hbar$, we obtain
\begin{eqnarray}
\langle\hat{\mathbf{R}}(t)\rangle =
e^{\frac{\hbar}{4}\mathbf{\mbox{\boldmath$\nabla$}}^2}\mathbf{r}(t) \cong
\mathbf{r}(t)+\frac{\hbar}{4}\mathbf{\mbox{\boldmath$\nabla$}}^2
\mathbf{r}(t),\label{O1h}
\end{eqnarray}
where $ \mbox{\boldmath$\nabla$}^2=\sum_{i=1}^{N}\nabla_i^2$ is the
$N$-dimensional Laplacian operator. Now, we formally define the Ehrenfest
time $t_E$ as being the instant at which the magnitude of the difference
between the quantum centroid and the corresponding classical vector in phase
space becomes equal to the magnitude of the initial classical vector.
Mathematically this condition can be expressed as follows:
\begin{eqnarray}
\frac{||\langle\hat{\mathbf{R}}(t_E)\rangle -\mathbf{r}(t_E)||}{||\mathbf{r}
(0)||}=1.
\label{tnk}
\end{eqnarray}
Finally, using Eq.(\ref{O1h}), we obtain the prescription for the
analytical calculation of the Ehrenfest time
\begin{eqnarray}
\frac{\hbar}{4}\frac{||\mbox{\boldmath$\nabla$}^2\mathbf{r}(t_E)||}{||
\mathbf{r}(0)||}=1.
\label{te}
\end{eqnarray}
It is remarkable that we do {\em not} need to solve the quantum
equations of motion to find $t_{E}$. In what follows, we will
calculate the above defined break time for the system of $N$
coupled non-linear oscillators.
%
%%%%%%%%%%%%%%%%%%%%%%%%%%%%%%%%%%%
\subsection{Application to a Non-Linear System}
%%%%%%%%%%%%%%%%%%%%%%%%%%%%%%%%%%%
%
Consider the following classical Hamiltonian
\begin{eqnarray}
\mathcal{H}=\sum\limits_{i=1}^{N}\omega_i \left(\frac{p_i^2+q_i^2}{2}\right)
+ g\left[\sum\limits_{i=1}^{N}\left(\frac{p_i^2+q_i^2}{2}\right) \right]^k,
\end{eqnarray}
where $k \ge 1$ is integer and $g$ is the only coupling constant of
the system, and from which we define the characteristic classical action
 $\Lambda=\sum\limits_{i=1}^{N}\left(\frac{p_i^2+q_i^2}{2}
\right)$ \cite{remark}. The equations of motion can be solved by 
noticing that $\Lambda$ itself is a constant of motion. The result reads
%%%%%%%%%%%%%%%%%%%%%%%%%%%%%%%%%%%%%%%%%%%%%%%%%%%%%%%%%%%%
\begin{figure}[ht]
\centerline{\includegraphics[scale=0.4,angle=-90]{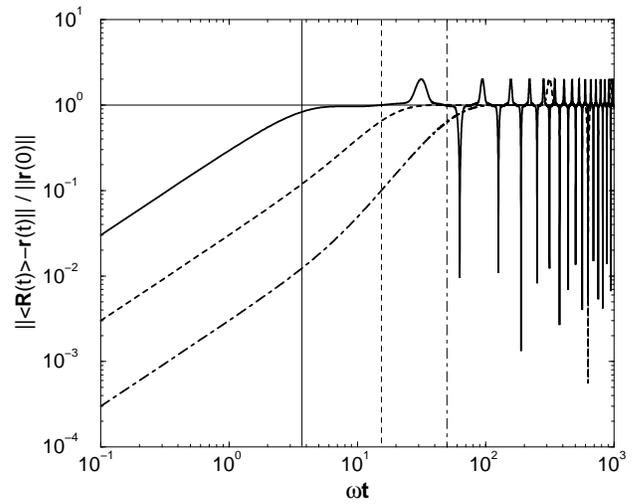}}
\caption{\small Departure between quantum and classical centroids
as a function of the dimensionless parameter $\omega t$ for the
bi-dimensional quartic oscillator $(N=2,k=2)$. These calculations
were analytically performed with $q_i=p_i=1$ ($\Lambda=2$),
$\omega_i=\omega=1$, $g=0.1$ and $h=1$ (solid line), $\hbar=0.1$
(dashed line) and $\hbar=0.01$ (dot-dashed line). The Ehrenfest time
scale (\ref{tnk}) for each value of $\hbar$ is represented by a vertical 
line in the corresponding style.} \label{fig:deltaR}
\end{figure}
%%%%%%%%%%%%%%%%%%%%%%%%%%%%%%%%%%%%%%%%%%%%%%%%%%%%%%%%%%%%%
\begin{eqnarray}
\left[\begin{array}{c} q_i(t) \\ \\ p_i(t)  \end{array}
\right]=\left[\begin{array}{cc} \cos{\Theta_i(t)} & \sin{\Theta_i(t)} \\ \\
-\sin{\Theta_i(t)} & \cos{\Theta_i(t)}
\end{array} \right]\left[\begin{array}{c} q_i \\ \\ p_i
\end{array} \right], \label{matrix}
\end{eqnarray}
where $(q_i,p_i)$ are the initial conditions of the $i$-th oscillator and
$\Theta_i(t)=\omega_i t+ g t k \Lambda^{k-1}$ is a rotation angle in
 phase space. Finally, using
(\ref{te}) and (\ref{matrix}), we get for the Ehrenfest time of this
system
\begin{eqnarray}
t_{E}&\cong
&\frac{1}{k(k-1)}\left[\frac{1}{g\Lambda^{k-1}}\left(\frac{2\Lambda}{\hbar}
\right)^{0.5}\right] \left(1-\frac{\hbar k²}{8\Lambda}
\right),\label{tnk}
\end{eqnarray}
where we kept only the first order terms in $\hbar$. The cases
$(N=1,k=2)$ and $(N=2,k=2)$ can be seen to reproduce respectively
the results calculated in ref.\cite{berman} and
\cite{carolina}. Trivial limits are also contemplated by the
above result, namely, the cases of an harmonic system ($g=0$ or
$k=1$), for which $t_E\to\infty$.
Moreover, our result is in accordance with some conjectures 
predicting the general form
$\frac{1}{\Omega}\left(\frac{S}{\hbar}\right)^{\delta}$ for the
break time of classically integrable systems \cite{zurek}, where
$\Omega= \frac{1}{k(k-1)g\Lambda^{k-1}}$ and $S\cong2 \Lambda$ are
respectively the typical frequency and classical action of the
system in consideration.
We illustrate in figure \ref{fig:deltaR} the Ehrenfest scale predicted 
by the expression (\ref{tnk}) for the case $(N=2,k=2)$, where we can 
see an algebraic ($\sim t²$) short time departure. We also note that 
the first order correction  to the Ehrenfest time scale seems to indicate 
a more appropriated parameter to measure 
the classicallity in this system:  
$\frac{\hbar k^2}{8\Lambda}\ll 1$.
%%%%%%%%%%%%%%%%%%%%%%%%%%%%%%%%%%%%%%%%%
\section{Conclusion}
%%%%%%%%%%%%%%%%%%%%%%%%%%%%%%%%%%%%%%%%%
In this paper we proposed a scheme for an
ordered symmetric quantization based on the action of certain 
differential operators. As an application, we quantized a classically time
evolved dynamical functions of the canonical variables in phase space, 
and showed that such a procedure is adequate during the Ehrenfest time 
scale for separable coherent initial states. This allowed us to propose a 
formal definition for the Ehrenfest time in terms of the phase space 
Laplacian operator acting on the classical solutions of the equations of 
motion. We performed an explicit calculation for a system of $N$ coupled 
nonlinear ($2k$-th order) oscillators and calculated the Ehrenfest time 
for general $N \ge 1$ and $k\ge 1$. The results are shown to agree with the 
results known in the literature for some particular cases.
\begin{acknowledgments}
It is a pleasure to acknowledge M.C. Nemes and A.C. Oliveira for many
helpful discussions, and Conselho Nacional de Desenvolvimento
Cient\'{\i}fico e Tecnol\'ogico (CNPq) (Contracts No.300651/85-6,,
No.146010/99-0) and Funda\c{c}\~ao de Amparo \`a Pesquisa de S\~ao Paulo
(FAPESP) (Contract No.98/13617-4) for financial support.
\end{acknowledgments}
%%%%%%%%%%%%%%%%%%%%%%%%%%%%%%%%%%%%%%%%%%%


\begin{thebibliography}{20}
\bibitem{remark} Notice that here we have an extra factor of $\frac{1}{2}$ 
as compared to the reference \cite{sanz}.
\end{thebibliography}
\end{document}